\documentclass[aps,amssymb,amsmath,reprint]{revtex4-1}

\usepackage{graphicx}  % needed for figures
\usepackage{dcolumn}   % needed for some tables
\usepackage{physics}
\usepackage{natbib}
\usepackage{color}
\usepackage{appendix}
\usepackage{ulem}
\usepackage{float}
\usepackage{changes}
\definecolor{orange}{rgb}{0.8, 0.3, 0}

\definecolor{blueviolet}{rgb}{0.2, 0.2, 0.6}
\usepackage[pdftex, % FOR PDFLATEX ONLY
bookmarks=true, % Bookmark bar
colorlinks=true,
allcolors=blueviolet,
pdfstartview={FitH}, % FitBH
]{hyperref}
\begin{document}

%\title{Improving loss and decoherence in superconducting circuits  by surface treatment of silicon substrates: Insights from electron spin resonance}
\title{Loss and decoherence in superconducting circuits on silicon: Insights from electron spin resonance}
\author{Aditya.~Jayaraman$^{1}$}
\email{adityaja@chalmers.se}
\author{Andrey.~V.~Danilov$^1$}
\author{Jonas~Bylander$^{1}$}
\author{Sergey.~E.~Kubatkin$^1$}
\affiliation{$^1$Department of Microtechnology and Nanoscience MC2, Chalmers University of Technology, SE-41296 Goteborg, Sweden}

%Abstract #######################################%
\begin{abstract} %% should be less than 600 characters incl space
Solid-state devices used for quantum computation and quantum sensing applications are adversely affected by loss and noise caused by spurious, charged two-level systems (TLS) and stray paramagnetic spins. These two sources of noise are interconnected, exacerbating the impact on circuit performance. We use an on-chip electron spin resonance (ESR) technique, with niobium nitride (NbN) superconducting resonators, to study surface spins on silicon and the effect of post-fabrication surface treatments. We identify two distinct spin species that are characterized by different spin-relaxation times %, which are also saturated at different microwave power levels 
and respond selectively to various surface treatments (annealing and hydrofluoric acid). Only one of the two spin species has a significant impact on the TLS-limited resonator quality factor at low-power (near single-photon) excitation.  We observe a 3-to-5-fold reduction in the total density of spins after surface treatments, and demonstrate the efficacy of ESR spectroscopy in developing strategies to mitigate loss and decoherence in quantum systems.
\end{abstract}

\maketitle
%%##################### Introduction #############################%%

%In recent years, the field of quantum computation has witnessed significant progress, while superconducting quantum circuits have emerged as a promising platform for the realization of quantum information processing ~\cite{Devoret2013,Martinis2020}. However, the performance of these circuits is limited by the detrimental effects of noise and decoherence~\cite{Muller2019,Paladino2014,Burnett2014}
%Although the scalability of superconducting qubit architectures, based on standard lithographic technology, is decisively advantageous, the incorporation of dielectric substrate inadvertently serves as the primary sources of noise and decoherence~\cite{Muller2019}. It is widely accepted that the substrate and substrate-superconductor interfaces host material defects which manifest as spurious two-level systems (TLS) and play a dominant role in fidelity loss of gate operations in superconducting circuits ~\cite{Paladino2014,Wang2015,Macha2010,Gambetta2017,Osman2023,Verjauw2021,Lisenfeld2019}. Despite the decennial efforts~\cite{Muller2019}, the precise material nature of these defects remains elusive, and mitigation strategies have largely been empirical.

%JB suggestion:
Contemporary solid-state devices for quantum computation and quantum sensing suffer from energy loss and noise originating from materials defects~\cite{Paladino2014,Burnett2014,Muller2019}. Predominantly located in dielectrics within the devices, these fluctuators cause quantum decoherence, which prevents reaching the performance level necessary for meaningful applications~\cite{Muller2019}. For superconducting devices---a leading platform for quantum computation---it is widely accepted that all interfaces (substrate--air, substrate--metal, and metal--air) host spurious defects which manifest as two-level systems (TLS)~\cite{Paladino2014,Wang2015,Macha2010,Gambetta2017,Osman2023,Verjauw2021,Lisenfeld2019}. Despite decades of research on TLS, their microscopic nature has remained elusive, and mitigation strategies have been largely empirical: to remedy this, we seek an improved understanding of the origin of these defects and informed strategies for their removal during or after device fabrication. 

%Silicon and sapphire are two commonly used substrates for superconducting quantum devices. Electron-spin resonance methods have recently been used to study TLS on sapphire: the two dominant species were identified as atomic hydrogen and $g = 2$ spins (where the presence of hydrogen is reduced by annealing). Here we perform such a study on silicon, the most popular substrate for high-coherence superconducting qubits, and report differences and augmented insight into these microscopic defects.

 \begin{figure}[b]
  \includegraphics[width=1\linewidth]{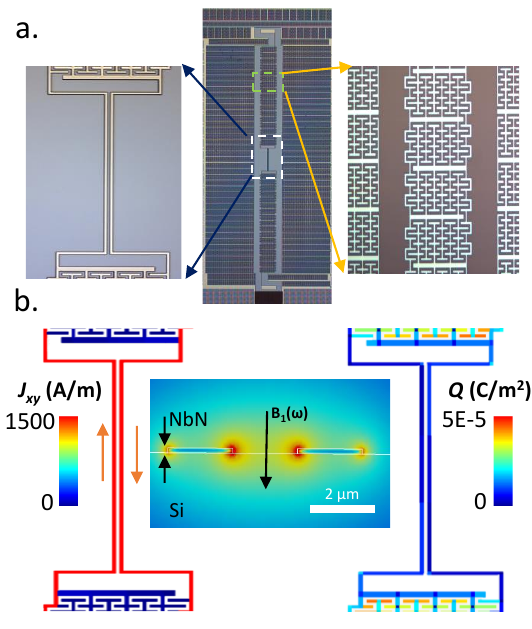} 
  \caption{\textbf{a.} Optical image of an NbN resonator with narrow striplines at its center. \textbf{b.} Sonnet simulation showing microwave current density ($J_{xy}$, left) and charge density ($Q$, right) distribution at the current anti-node. The microwave magnetic field ($B_1$) is primarily confined to a small area  between the striplines (center).}
  \label{fig1}
\end{figure}

TLS can exhibit an electric dipole moment, resulting in charge noise and energy loss in superconducting resonators and qubits~\cite{DeGraaf2018,Faoro2012,Burnett2014,Burnett2019,martinis2005,Lucas2022}.
%, or a spin, which is a source of  magnetic-flux noise~\cite{Kumar2016,Anton2013,Quintana2017, Choi2009}. 
Noise with a $1/f$-type spectrum is comprehensively explained by the generalized tunneling model (GTM), which considers the dipole-dipole interaction between the TLS~\cite{Faoro2015}. 
%Similarly, $1/f$ noise observed in flux qubits and superconducting quantum interference devices (SQUIDs) occurs as a result of the coupling to stray spins. 
%Several types of spin defects, such as surface dangling bonds~\cite{desousa2007}, adsorbed molecules~\cite{DeGraaf2017,wang2015a,Adelstein2017}, and intrinsic nuclear spins~\cite{Laforest2015}, have been proposed as potential sources of flux noise. 
%While magnetic spins are not directly addressed by the GTM, typically, surface defects can possess both charge and spin, with a potentially more intricate effect on quantum devices.  
%
The GTM distinguishes between two different types of fluctuators: coherent, quantum two-level systems (cTLS) and incoherent, classical two-level fluctuators (TLF). 
While resonant cTLS are responsible for energy loss in resonators, the slowly fluctuating TLF impart an energy drift on the resonant cTLS, resulting in charge (dielectric) noise. 
%Understanding the nature of both cTLS and TLF is essential for a comprehensive grasp of the decoherence pathways in quantum systems. 

Similarly, TLS can possess spin, which is a source of  magnetic-flux noise~\cite{Kumar2016,Anton2013,Quintana2017, Choi2009}; 
the $1/f$ noise observed in flux qubits and superconducting quantum interference devices (SQUIDs) occurs as a result of the coupling to stray spins. 
Several types of spin defects, such as surface dangling bonds~\cite{desousa2007}, adsorbed molecules~\cite{DeGraaf2017,wang2015a,Adelstein2017}, and intrinsic nuclear spins~\cite{Laforest2015}, have been proposed as potential sources of flux noise. 
While magnetic spins are not directly addressed by the GTM, typically, surface defects can possess both charge and spin, with a potentially more intricate effect on quantum devices.

Silicon and sapphire are substrate materials commonly used for superconducting quantum devices. 
%Electron-spin resonance methods have recently been used to study TLS on sapphire: the two dominant species were identified as atomic hydrogen and $g = 2$ spins (where the presence of hydrogen is reduced by annealing). 
Recent studies of sapphire suggest that paramagnetic surface spins act as TLF, and their removal leads to reduced charge noise in resonators, underscoring the importance of spins in the context of quantum circuit performance~\cite{DeGraaf2018,Un2022}. 
%However, no experimental data revealing the microscopic properties of TLS is available to date. 
%It is important to strive to understand the microscopic nature of both cTLS and TLF in order to identify and mitigate the sources of decoherence in these systems. 

\begin{figure}[t]
  \includegraphics[width=1\linewidth]{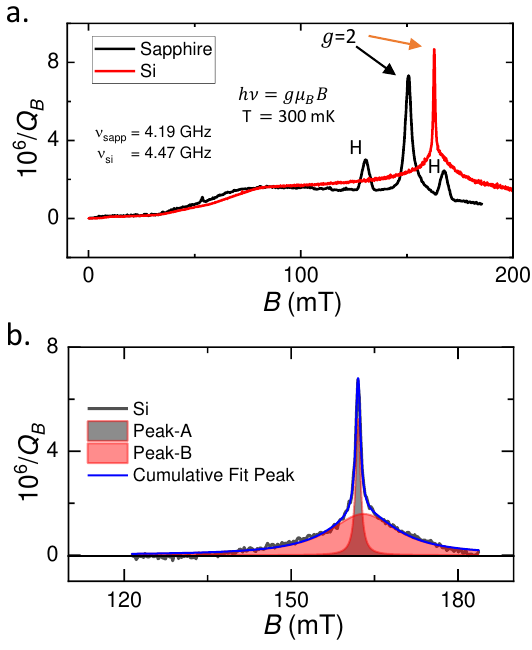} 
  \caption{\textbf{a.} ESR spectra of surface spins on silicon (red) and sapphire (black) measured at $T= 300$~mK with an NbN resonator~\cite{DeGraaf2017}. The salient features, including the hydrogen satellite peaks (in the sapphire spectrum) and the central $g=2$ peak, are labelled. \textbf{b.} $g=2$ region of the spectrum, showing the decomposition into two Lorentzian peaks (A and B) with different linewidths.}
  \label{fig2}
\end{figure}

In this work, we employ an on-chip electron-spin resonance (ESR) technique to identify the composition of dilute spins on intrinsic silicon (Si), the most popular substrate for high-coherence superconducting qubits. Our ESR spectra reveal two distinct spin families residing on the silicon surface. Further measurements to determine the corresponding dissipative losses allow us to associate one family with coherent (cTLS) and the other with non-coherent (TLF) fluctuators. We demonstrate that moderate annealing of the chip selectively removes TLF, while treatment with hydrofluoric acid (HF)  removes spins associated with cTLS; these two post-fabrication treatments therefore target different contributing sources to the decoherence pathway. 
%We observe a significant reduction in spin density after surface treatments and show direct correlation between the energy loss in resonators and the ESR spectra, revealing important insights on the nature of surface TLS on silicon-based quantum circuits. 
%A similar approach was used in the study of surface spins on sapphire, where we observed the presence of atomic hydrogen adsorbed on the surface, which was removed upon annealing~\cite{DeGraaf2017}. 

%%%%%%Results%%%%%%
The resonators used in this work are descendants of a design reported by de~Graaf~\cite{DeGraaf2014}, shown in Fig.~\ref{fig1}a, with a `quasi-fractal' geometry optimized for high sensitivity to surface spins in ESR measurements~\cite{DeGraaf2012,DeGraaf2017,Mahashabde2020,Keyser2020} (rather than to obtain the highest possible Q-factor). 
The circuits were etched out of a 120 nm thick NbN film sputtered onto an intrinsic silicon substrate and patterned using electron beam lithography, see Methods~\cite{supp}.

The resonator's fundamental harmonic is a half-wavelength ($\lambda/2$) mode, which has a current anti-node at the center. The anti-node region comprises a pair of superconducting strips measuring $2~\mu$m in width, with a spacing of $2~\mu$m. The magnetic fields generated by the strips interfere constructively between the strips and destructively elsewhere, so that the microwave field is effectively confined to a micrometer-scale volume, as illustrated in the inset of Fig.~\ref{fig1}b. Consequently, the resonator couples primarily to the surface spins in this area, and its surface spin resolution is about $10^{5}-10^{6}$ spins/mm$^{2}$. 
%A similar approach was used in the study of surface spins on sapphire, where we observed the presence of atomic hydrogen adsorbed on the surface, which was removed upon annealing~\cite{DeGraaf2017}. 

%Therefore, most of the spins that couple to the resonator are mainly present at the anti-node.  The anti-node comprises a pair of superconducting strips which are $2~\mu$m wide and has a pitch of $2~\mu$m. The microwave field is therefore confined in the micron scaled volume as shown in the inset in Fig.\ref{fig1}~(a). This ensures that the resonator robustly couples to and hence predominantly examines the spins on the surface. The strong coupling to the spins and high Q-factors compared to conventional ESR spectrometers make them ideal candidates for investigating dilute surface spins on the surface of the silicon. Further details about the fabrication and the measurements can be found in the supplementary material~\cite{}.

A typical ESR spectrum measured using an NbN resonator on silicon is presented in Fig.~\ref{fig2}a (red). The plot represents the field-dependent part of the total energy losses in the resonator, $Q_{B}^{-1}(B)=Q_{i}^{-1}(B)-Q_{i}^{-1}(0)$, where the internal Q-factor ($Q_i$) is extracted from the S$_{21}$ resonance data as a function of applied parallel magnetic field ($B$). The spectrum measured on silicon (red) is in contrast with that on sapphire, see Fig.~\ref{fig2}a (black). On silicon, we observe a pronounced sharp peak corresponding to a spin ensemble with a g-factor $g=2$. The peak's linewidth (1.2~mT), however, is less than that on sapphire (5~mT), implying the presence of more coherent $g=2$ spins on silicon. 
Interestingly, both the silicon and sapphire peaks are superimposed on a broad background pedestal with an onset at approximately 50~mT.
Moreover, on sapphire, the spectrum reveals the presence of atomic hydrogen on the surface, as evidenced by the presence of additional satellite peaks with a separation ($\Delta B$) corresponding to 1.42~GHz~\cite{DeGraaf2017,DeGraaf2018,Un2022}. The absence of satellite peaks on silicon suggests that atomic hydrogen is not adsorbed on its surface (and also indicates that it is not adsorbed on the surface of NbN in both devices). This could be attributed to the lower adsorption energy of hydrogen on silicon relative to sapphire, although hydrogen is known to passivate surface defects on silicon~\cite{Brower1990}. 

While the central $g=2$ peak in the sapphire spectrum can be fit with a single Lorentzian~\cite{DeGraaf2017}, the silicon spectrum cannot---instead it requires two peaks with different linewidths, see Fig.~\ref{fig2}b. This indicates the presence of two distinct spin communities which both possess the same g-factor. The linewidth of the sharper peak (A) is $\gamma_2/2\pi=g\mu_B\Delta B/h= 33$~MHz, whereas that of the wider peak (B) is $670$~MHz (here $\mu_B$ is the Bohr magneton and $h$ is Planck's constant). Assuming that the peaks are homogeneously broadened, as indicated by faithful Lorentzian fits, the spin relaxation times ($T_{2e}$) can be directly estimated from the linewidths and are found to be $30$~ns for Peak A and $1.5$~ns for Peak B.

\begin{figure}[t]
  \includegraphics[width=1\linewidth]{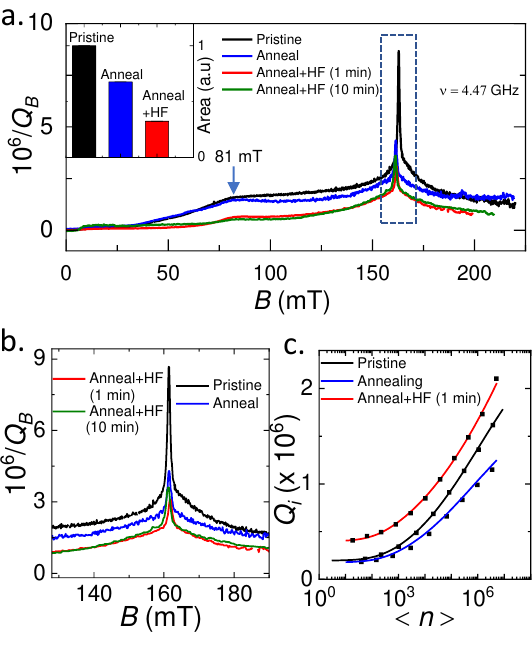} 
  \caption{\textbf{a.} ESR spectra of an NbN-on-Si resonator taken before and after surface treatments. An arrow marks the field corresponding to half of the $g=2$ resonance. Post-annealing, the $g=2$ peak shifts slightly due to a reduction in the resonator's frequency caused by changes in the NbN film's kinetic inductance. \textbf{Inset:} The total area of the ESR spectra before and after treatments.  \textbf{b.} Effect of treatments on the central $g=2$ peak. The minor shifts in panel \textbf{a.} are adjusted for clarity. \textbf{c.} Internal Q-factor ($Q_i$) as a function of average photon number before and after treatments. Solid lines represent fits to the TLS model in Eq.~\ref{e1}.}
  \label{fig4}
\end{figure}

\begin{figure*}[t]
  \includegraphics[width=1\linewidth]{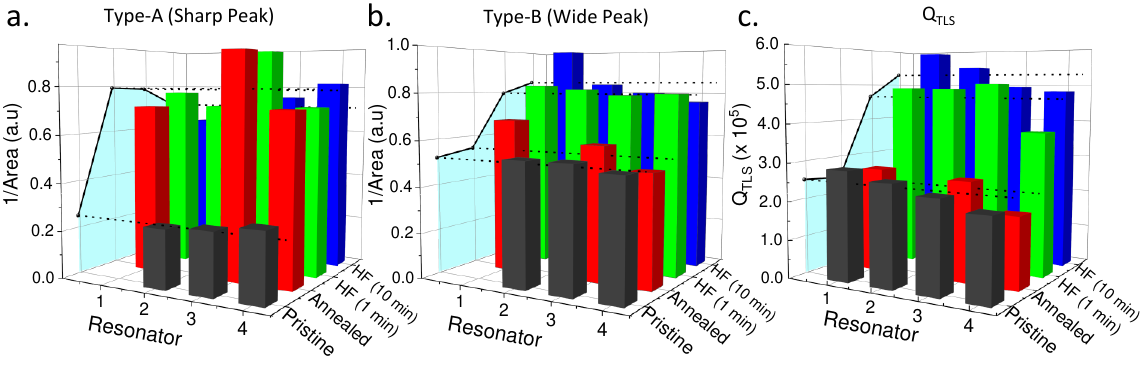} 
  \caption{Effect of annealing and HF immersion treatments on four resonators with resonant frequencies 3.88, 4.09, 4.47, and 4.90~GHz. 
  \textbf{a--b.} Inverse area of the sharper Peak A and the wider Peak B, respectively. \textbf{c.} TLS-limited resonator Q-factors, $Q_{TLS}$. The average values of $Q_{TLS}$ and areas across all resonators are projected onto the left coordinate plane (shaded in cyan). (Some resonators were not measurable due to interference with spurious ground-plane resonances.)}
\label{fig5}
\end{figure*}

We will now explore the impact of surface treatments on the two types of identified spin groups (from here on referred to as `type-A' and `type-B' spins, corresponding to Peaks A and B). Two specific surface treatments were carried out on the silicon sample: annealing at 300$^{\circ}$C for 30 minutes in vacuum and submersion in a 5$\%$ HF solution; the latter was done for either 1~min or 10~min, on different samples. %, with subsequently negligible observed difference in the spectra. 
Figure~\ref{fig4}a illustrates the ESR spectrum prior to treatment as well as after each treatment.
%A comparison of the annealed sample's spectrum to that of the untreated sample reveals 
We observe that annealing causes a notable decrease in the sharp $g=2$ peak (A) and a minor decrease in the background signal, see Fig.~\ref{fig4}b, whereas HF treatment causes a reduction of Peak B and a significant decline in the background signal, with the step at approximately 80 mT almost entirely disappearing; however, HF does not significantly further reduce Peak A. 
We note that the linewidths of the peaks do not change, signifying a reduced density of surface spins without any affect on the relaxation timescales~\cite{supp}. 
%The ESR spectrum after 10 minutes of HF immersion exhibits negligible variation compared to the spectrum after 1 minute of immersion. 
By calculating the integrals of the areas under the ESR spectra for each case, we can directly compare these spin densities (inset in Fig.~\ref{fig4}a). 
Furthermore, the separate integrated areas of Peak A and B are shown in Fig.~\ref{fig5}a and b, respectively, calculated for four resonators when pristine and after annealing and HF-immersion.
Following both treatments, we observe a significant 3-to-5-fold reduction in total spin density.

While annealing and HF have different effects on the concentrations of the two types of spins, we also notice different impacts on the resonator Q-factors, as shown in Fig.~\ref{fig4}c. We fit the power dependence of $Q_i$ of the resonator according to the model for interacting TLS~\cite{Lucas2022},
\begin{equation}
    \frac{1}{Q_i(\langle n \rangle)} = \delta_0 + \frac{1}{Q_{TLS}}\left(1+\frac{\langle n \rangle}{n_c}\right)^{-\beta}
    \label{e1}
\end{equation}
where $\delta_0$ represents power-independent, non-TLS-related dissipation, $\langle n\rangle$ is circulating power in the resonator, expressed as the mean photon population, and  $n_c$ is the average number of photons required to saturate a single TLS. The power exponent $\beta$ indicates the rate at which TLS saturate with power: it is 0.5 for non-interacting TLS but has a smaller value in the more general case of an interacting-TLS model, and the value we extract from our data ($\sim\! 0.2$) is consistent with that reported in literature~\cite{Faoro2015,Lucas2022,DeGraaf2018}.
 
Following annealing, we observe a minimal change in the TLS-limited Q-factor ($Q_{TLS}$) at low power. In contrast, submersion in HF significantly improves $Q_{TLS}$, doubling its value.
Figure~\ref{fig5}c presents the distribution of $Q_{TLS}$ values before and after each treatment across four different resonators, substantiating a decrease in resonator losses upon HF immersion, due to a two-fold decrease of Peak B. 
%This distribution is compared against the areas of Peak A (Fig.~\ref{fig5}a) and Peak B (Fig.~\ref{fig5}b), correlating to the density of each spin species on the surface. 
Extending the HF immersion to 10 minutes does not result in a reduction in spin density or a significant improvement in $Q_{TLS}$.
However, it is intriguing that the annealing-induced reduction of spins corresponding to the sharper Peak A does not improve the Q-factor. 

The microscopic nature of the radicals that contribute to $g=2$ peaks remains uncertain, as it is a characteristic of multiple different spin systems with free-electron-like g-factors. However, the surface treatments can provide insights about the nature of the two spin groups. Numerous plausible candidates could give rise to these spins---for instance, the (001) surface of silicon is known to harbor numerous paramagnetic defects with g-factors close to 2~\cite{Lenahan1998}. These could manifest as dangling bonds in the Si/SiO$_2$ interface ($P_b$ centers)~\cite{caplan2008,poindexter1984}, oxide trapping centers ~\cite{Lenahan1998}, defects in oxides of niobium~\cite{Verjauw2021,Altoe2022}, among other possibilities. Previous reports indicate that $P_{b0}$ centers (i.e. dangling bonds with trivalent silicon) dominate the interface traps at the (100) Si/SiO$_2$ boundary~\cite{Ohdomari1981,poindexter1984}. Annealing the sample results in the migration and rearrangement of these defects and therefore leads to surface passivation, reducing the ESR signal. Such a reduction in amorphous silicon was observed previously~\cite{Ohdomari1981} and corroborates our results. While other sources of $g=2$ spins are not ruled out, our findings from surface treatments suggest that these dangling-bond defects are primarily accountable for the sharper Peak A. 

The existence of the wider Peak B, which responds differently to surface treatments, is more intriguing. The reduction in type-B spins after annealing is modest, but after HF immersion it is significant, which suggests that these spins are likely located within the oxide layer on silicon.
Recent studies reveal that niobium oxides are removed at a much slower pace (2.2 pm/s) than silicon oxides (1.8 nm/s)~\cite{Altoe2022}. Since the type-B spins show only a minor or no reduction after 10 minutes in HF, they are unlikely to be found within the niobium oxides. Furthermore, the absence of a similar broad peak in the NbN-on-sapphire spectrum leads us to discount the possibility of spins located exclusively at the NbN/air interface.

We note another interesting observation regarding the background signal. While the $g=2$ peak appears around 162 mT for the 4.47 GHz resonator, the pedestal peaks at precisely half the field, i.e.\@ at 81~mT (Fig.~\ref{fig4}a). As shown in the supplementary (Fig.~S4), at elevated microwave powers, once the background signal is effectively suppressed, one can resolve a Gaussian peak at around 80~mT. This observation suggests that the background signal partially originates from spins with $S=1$. This may be attributed to spin-spin interactions and clustering of spins, resulting in the formation of spin-triplet centers. These mechanisms are considered potential contributors to $1/f$ flux noise~\cite{stevenson1984,Atalaya2014}. 

The impact of treatments on $Q_{TLS}$ offers additional insights into the nature of spin ensembles. Within the framework of the generalized tunneling model, cTLS that are resonant with the resonator engage in energy exchange and dissipation into the surrounding environment, leading to energy loss in resonators. On the other hand, the interaction of coherent TLS with a bath of thermally activated TLF leads to spectral drift of the resonant frequency, a key factor contributing to its temporal fluctuations~\cite{Faoro2015,DeGraaf2018}. 

The negligible reduction in loss ($1/Q_i$) observed upon annealing, coupled with a dramatic decrease in the density of the type-A spins suggests that, while these spins are removed, they do not affect the density of the cTLS. Most of the commonly discussed spin radicals in literature, such as dangling bonds, adatoms, defects, and vacancies, are known to exhibit both spin and charge (or electric dipole) characteristics~\cite{Ohdomari1981,Lenahan1998}. We therefore conclude that charges/dipoles associated with type-A spins function as classical TLF. This corroborates our previous findings in sapphire, where desorption of the spins does not result in a significant reduction of loss but does lead to a ten-fold reduction of the noise~\cite{DeGraaf2018}. Immersion in HF, however, reduces the TLS losses by a factor of two, and concurrently reduces the density of type-B spins by the same factor (Fig.~\ref{fig5}). This implies that type-B spins are associated with coherent two-level systems and are of a distinct nature compared to the type-A spins. Hence, despite having similar g-factors, the two spin species have different origins and different influence on decoherence.

We also highlight that type-B spins exhibit a significantly wider peak (smaller T$_{2e}$) compared to type-A spins. We speculate that this difference may be attributed to the fact that TLS-related dissipation comes from near-resonant charge fluctations, and under ESR resonance condition the spins of those TLS also come into resonance with the microwave field. This concurrent near-resonance of both the spins and associated charges of type-B radicals should coherently couple spins and charges through photons. To our knowledge, this exotic interaction has not been explored in the existing literature, particularly in relation to decoherence mechanisms, and we hope our findings will stimulate further work in this direction.

In conclusion, our study sheds light on the presence and impact of surface spins on silicon substrates in the context of superconducting quantum systems. Using an on-chip ESR technique, we have identified two distinct groups of spins, each demonstrating a unique response to power variations and surface treatments. We have reduced the density of paramagnetic spins by 3 to 5 times, which we anticipate will significantly decrease both flux and charge noise levels. %thereby helping mitigate decoherence in quantum systems.  
Our study categorizes `type-A' spins as incoherent two-level fluctuators, known contributors to charge noise and to energy shifts in coherent TLS. Conversely, `type-B' spins seem to be linked with coherent TLS on the silicon surface, directly influencing resonator loss. Therefore, a combination of ESR spectroscopy with measurements of energy loss has been instrumental in distinguishing between coherent and non-coherent fluctuators in silicon-based quantum circuits. By employing advanced spectroscopic techniques such as electron-nuclear double resonance (ENDOR) and electron-electron double resonance (ELDOR), one can acquire intricate details about the elemental and structural composition of TLS/TLF hosts (\textit{cf} Ref.~\cite{Un2022}) thereby offering promising avenues for informed noise and decoherence mitigation in quantum systems. 
%Our findings offer crucial insights into the management and understanding of surface spins, hinting at a potential interplay between stray magnetic and electric dipoles on silicon, thereby paving the way for improved performance and reliability of superconducting quantum circuits.  

\section*{Acknowledgments}
We are grateful for discussions with Anita Fadavi Roudsari, Amr Osman, and Janka Biznárová.
This work was performed in part at Myfab Chalmers. We acknowledge support from the Knut and Alice Wallenberg foundation via the Wallenberg Center for Quantum Technology (WACQT) and from the EU Flagship on Quantum Technology HORIZON-CL4-2022-QUANTUM-01-SGA project 101113946 OpeSuperQPlus100.

%##########  Bibliography  ######################
%
\newpage
%%%FIG4_other option%%%%
%\begin{figure*}[t]
%  \includegraphics[width=1\linewidth]{Fig5b.pdf} 
%  \caption{Effect of annealing and HF immersion treatments on four resonators with resonant frequencies %3.88, 4.09, 4.47, and 4.90~GHz. 
%  \textbf{a--b.} Inverse area of the sharper Peak A and the wider Peak B, respectively. \textbf{c.} TLS-%limited resonator Q-factors, $Q_{TLS}$. (Some resonators were not measurable due to interference with %spurious ground-plane resonances.)}
%\label{fig5b}
%\end{figure*}

%\begin{figure*}[t]
%  \includegraphics[width=1\linewidth]{Fig5c.pdf} 
%  \caption{Effect of annealing and HF immersion treatments on four resonators with resonant frequencies 3.88, 4.09, 4.47, and 4.90~GHz. 
%  \textbf{a--b.} Areas of the sharper Peak A and the wider Peak B, respectively. \textbf{c.} Disscipative losses (1/$Q_{TLS}$) at lower power levels (near single-photon excitation). Some resonators were not measurable due to interference with spurious ground-plane resonances.}
%\label{fig5c}
%\end{figure*}
\clearpage
\onecolumngrid
\setcounter{figure}{0} % Reset the figure counter
\begin{widetext}
%\documentclass[aps,amssymb,amsmath,notitlepage]{revtex4-1} 

%\usepackage{graphicx}  % needed for figures
%\usepackage{dcolumn}   % needed for some tables
%\usepackage{physics}
%\usepackage{natbib}
%\usepackage{color}
%\usepackage{appendix}
%\usepackage{ulem}

%\definecolor{orange}{rgb}{0.8, 0.3, 0}

%\definecolor{blueviolet}{rgb}{0.2, 0.2, 0.6}
%\usepackage[pdftex, % FOR PDFLATEX ONLY
%bookmarks=true, % Bookmark bar
%colorlinks=true,
%allcolors=blueviolet,
%pdfstartview={FitH}, % FitBH
%]{hyperref}

\newpage

%\begin{document}

\title{Loss and decoherence in superconducting circuits on silicon: Insights from electron spin resonance}
% \author{Aditya.~Jayaraman$^{1}$}
% \email{adityaja@chalmers.se}
% \author{Andrey.~V.~Danilov$^1$}
% %\author{S.~E.~de~Graaf$^{1}$}
% \author{Jonas Bylander$^{1}$}
% \author{Sergey.~E.~Kubatkin$^1$}

%\affiliation{$^1$Department of Microtechnology and Nanoscience MC2, Chalmers University of Technology, SE-41296 Goteborg, Sweden}

\maketitle
\section*{Supplementary Information}

\section{Methods}

The resonators are fabricated on a high resistivity ($\rho > $ 10 k$\Omega$.cm) intrinsic silicon (100) substrate. The substrate was dipped in 2\% HF solution to remove native oxide and then subsequently rinsed with deionized water. The wafer was then immediately loaded in a metal sputter chamber. Before the deposition of NbN, the samples were annealed in-situ at 600$^\circ$C for 20 minutes, and a 2 nm seed layer was deposited. After being cooled to room temperature, an additional 120 nm of NbN was sputtered. The resonators were then patterned using electron beam lithography (UV60 resist: 33 $\mu$C/cm$^2$ dose, MF-CD-26 developer, DI water rinse) and were subsequently etched in an Ar/Cl$_2$ plasma. The fabrication procedure is similar to that in Ref.~\cite{DeGraaf2017,mahashabde2020}.

All experiments were conducted at a temperature of \(300 \, \text{mK}\) using a \(\text{He-3}\) single-shot refrigerator. After the initial measurements, the sample was subjected to a series of surface treatments. First, it was annealed in a vacuum at \(300^{\circ}\text{C}\) for \(30 \, \text{minutes}\) and subsequently re-measured. Second, it was immersed in a \(5\% \, \text{HF}\) solution for \(60 \,  \text{seconds}\) and then promptly reloaded into the cryostat within 45 to 60 minutes. This procedure was later repeated, but with the HF immersion time extended to \(10 \, \text{minutes}\). 

\section{Lorentzian Fits}
Fig.~\ref{figs1} shows the Lorentzian fits to the spectrum before and after surface treatment. It is evident from the figure that it results in a noticeable decrease in the amplitude of the peaks. 
However, treatments cause little to no effect on peak widths, suggesting that the $T_2$ relaxation time remains largely unchanged. Instead, this observation points to a reduction in the density of spins as the primary effect of the treatments.
\renewcommand{\thefigure}{S\arabic{figure}}

\begin{figure}[h!]
  \includegraphics[width=1\linewidth]{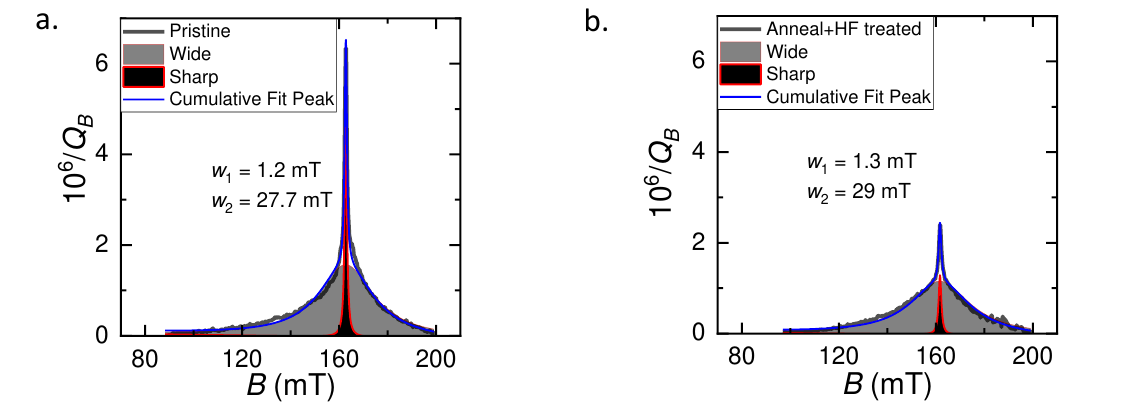} 
  \caption{Lorentzian fits to the central $g=2$ peak \textbf{a.} before, and \textbf{b.} after surface treatments.}
  \label{figs1}
  \end{figure}

\newpage
\section{Power dependence of ESR peaks}

\begin{figure}[h]
  \includegraphics[width=1\linewidth]{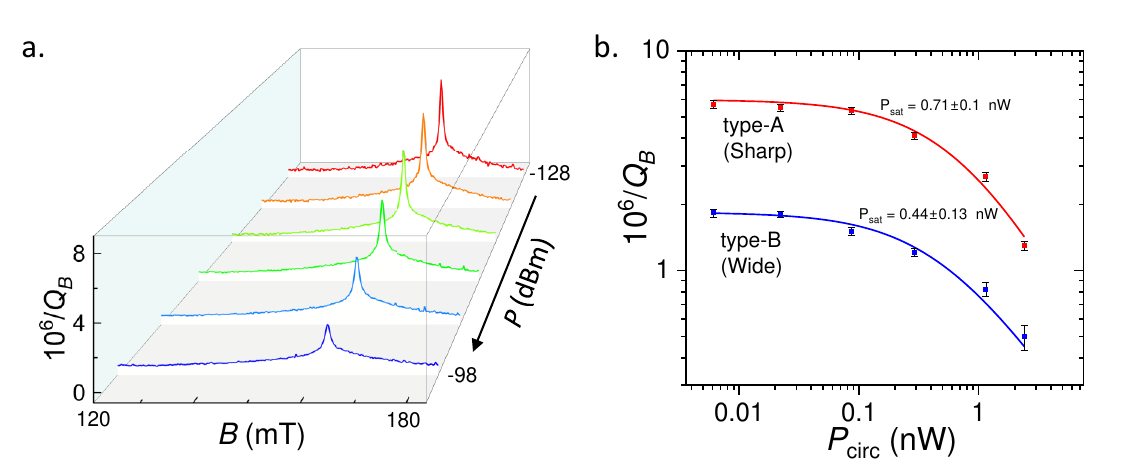} 
  \caption{\textbf{a.} ESR spectrum for various microwave drive powers at 300~mK. \textbf{b.} Inverse of the quality factor associated with energy dissipation into each individual spin system as a function of circulating power ($P_{circ}= 2Q^2P_{drive}/Q_{c}$), with fits to Eq.~\ref{eq:Qs}.}
  \label{figs2}
\end{figure}
The ESR spectra were measured as a function of microwave power as shown in Fig.~\ref{figs2}. With increasing power, a larger number of spins are driven out of equilibrium, which leads to diminished spin-induced dissipation, see Fig.~\ref{figs2}(a). By fitting two Lorentzian curves and decomposing the peak, we can extract the power dependence of spin-induced dissipation from individual species, 
%The dissipation follows the following equation~\cite{Haas1993}:

\begin{equation}\label{eq:Qs}
    \frac{1}{Q_B (P)} = \frac{1}{Q_{B0}}\left(\frac{1}{1+P/P_{sat}}\right)^{\epsilon}
\end{equation}
where the saturation power is $P_{\text{sat}} = 1/T_{1e} T_{2e} \gamma_{e}^{2} \alpha^{2}$, $\epsilon\sim 1$ is the inhomogeneity parameter describing how quickly spins saturate to increasing power~\cite{Haas1993},  $\gamma_{e}$ is the gyromagnetic ratio, and $\alpha = H/\sqrt{P_{0}}$ is the microwave power to microwave magnetic field conversion coefficient. An approximate value for our resonator, considering two parallel superconducting strips, is $\alpha= 0.21$~T/$\sqrt{\mathrm{W}}$~\cite{DeGraaf2017}.

The saturation power for type-A spins is $0.71 \pm 0.1$ nW, and for 
type-B spins it is $0.44 \pm 0.13$ ~nW. Given that Peak~B has a shorter relaxation timescale ($T_{2e}$), it is anticipated to relax at a higher power in comparison to the sharp peak. However, in our observations, both type-A and type-B spins saturate at approximately the same power, within the margins of error.  

In our work, all ESR spectra were obtained at power levels below the saturation threshold ($P < P_{\text{sat}}$), ensuring non-saturating conditions throughout the measurements.
%Nevertheless, this apparent inconsistency might be attributed to the coherent interaction between both the electrical and magnetic dipoles of type-B spins and the microwave field. Further research is necessary to fully understand this phenomenon.

%Although the complex nature of the resonator field distribution does not allow reliable estimations of spin coupling strength, and hence $T_{1e}$ for both spins, their ratio ($T_{1e}^{wide}/T_{1e}^{sharp}$) however, can be estimated assuming that the spins are identically distributed on the surface. Using $P_{sat}$ from Eq.~(1), we estimate the ratio to be $T_{1e}^{wide}/T_{1e}^{sharp} = 28$.  

\newpage
\section{Effect of HF treatment without annealing procedure}

\begin{figure}[h!]
  \includegraphics[width=1.0\linewidth]{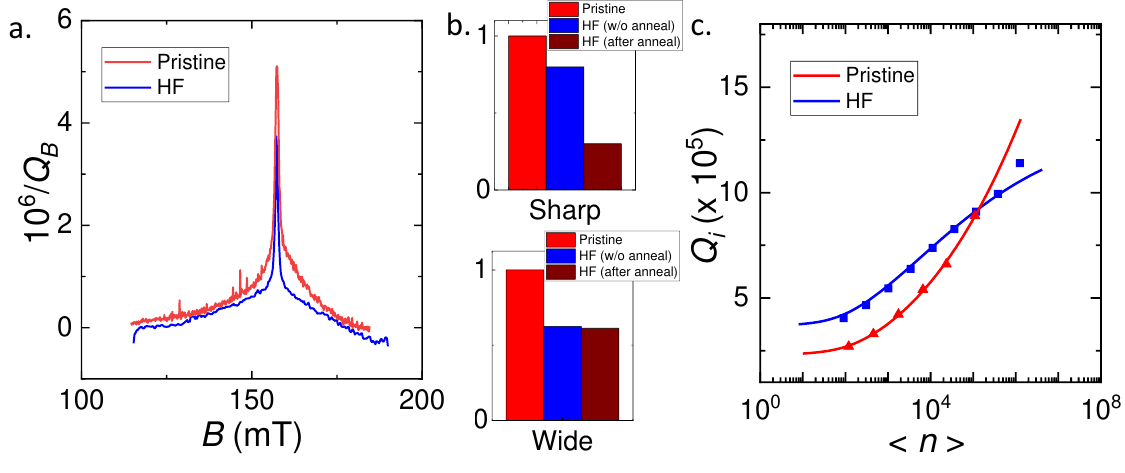} 
  \caption{\textbf{a.} ESR spectra of the 4.47-GHz resonator before and after HF treatment for 60 seconds. This sample was not subjected to annealing. \textbf{b.} Minor reduction in the normalized area of Peak A and two-fold reduction of Peak B after HF treatment. The area is compared to the HF-treated sample after annealing.  \textbf{c.} $Q_i$ vs.\@ $\langle n \rangle$, showing a corresponding two-fold improvement in $Q_\mathrm{TLS}$.}
  \label{figs3}
  \end{figure}
In the majority of samples, hydrofluoric acid (HF) treatment was carried out  after the annealing procedure. This resulted in enhancements in the Q factor and a decrease in the amplitude of the wide peak. To decouple the effects of annealing and HF treatment, we measured the ESR spectrum of a sample treated with HF for 60 seconds without preceding annealing (Fig.~\ref{figs3}). The results are consistent with our previous samples treated with HF after annealing: there is a minor reduction in the area of the sharp peak, a two-fold reduction in the area of the broad peak, and a corresponding two-fold improvement in Q$_{TLS}$. These findings suggest that the impacts of HF treatment are independent of those stemming from annealing. 

The high power (radiation-limited) loss in our resonator ($\delta_0$), however, does not show a consistent improvement with surface treatment. This can be attributed to different configurations of the wire bonds to the ground plane while rebonding after treatment. It is known that such configurations can inadvertently change the electromagnetic environment and introduce parasitic modes within the ground plane, resulting in additional loss channels that adversely impact the overall performance of the resonator at higher circulating powers.
\newpage
\section{ESR Background spectrum and S $=1$ peak}

\begin{figure}[h!]
  \includegraphics[width=1.0\linewidth]{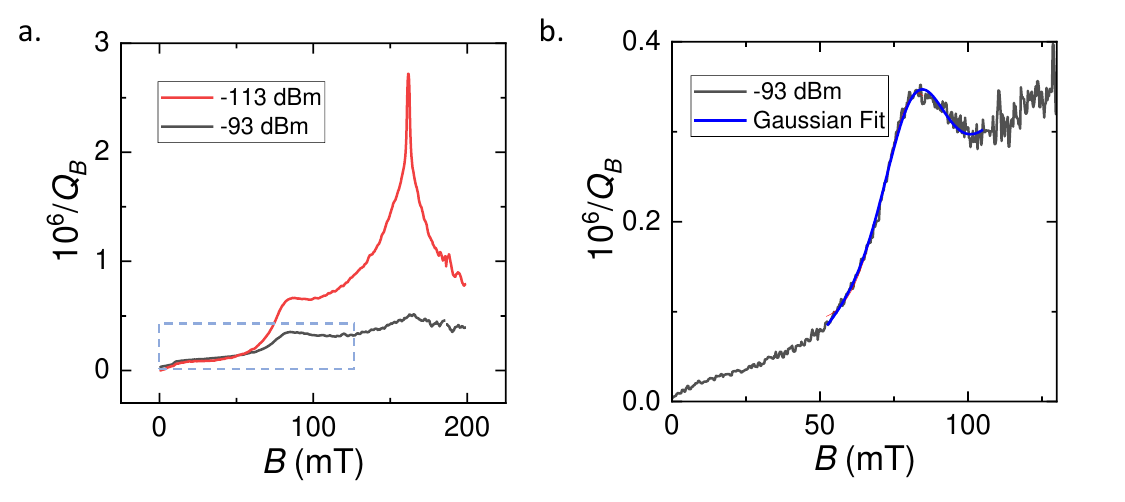} 
  \caption{\textbf{a.}  ESR spectrum of the sample following surface treatments, at both lower and higher microwave power levels. \newline\textbf{b.}  Enlarged view of the high-power spectrum, revealing the peak at half the field value of the $g=2$ peak (approx.\@ 80~mT), which has been distinctly resolved and fit with a Gaussian function.}
  \label{figs4}
  \end{figure}

Figure~\ref{figs4}a shows the ESR spectrum of the 4.47-GHz resonator after surface treatments at two different driving microwave powers. At higher power, both peaks corresponding to $g=2$ spins are effectively suppressed, along with a significant reduction in the background signal, as a result of spin saturation. Once the background noise is substantially reduced, a distinct peak becomes observable at approximately 80 mT, which is half the field strength typically associated with $g=2$ spins ($S = 1/2$). This particular peak is a hallmark of triplet-state systems with $S = 1$ and is believed to emerge from the interactions among electron spins and spin clustering~\cite{Stevenson1984,S1bertrand}. The resolved peak can be reliably fit with a Gaussian function with a linewidth of 20 mT. This suggests that the peak is in-homogeneously broadened, as illustrated in Fig.~\ref{figs4}b.   
 
%##########  Bibliography  ######################
%\bibliographystyle{}
%

%\end{document}
\end{widetext}

\end{document}